\documentclass[aps,prl,twocolumn,10pt,floatfix, superscriptaddress, showpacs]{revtex4-1}
\usepackage[pdftex]{graphicx}
 \usepackage{amsmath}
\usepackage[utf8x]{inputenc}
\usepackage[T1]{fontenc}
\usepackage{subfigure}
\usepackage{amssymb}
\usepackage{flushend}

 \usepackage{tikz}
\usetikzlibrary{decorations.markings}
\usepackage{amsfonts}
\usepackage{bm}
 \usepackage{amsmath} 
\usepackage{amsfonts} 
\usepackage{amssymb, mathrsfs}
\usepackage{braket}
\usepackage{graphicx} 

\usepackage{bbm}
\def\beq{\begin{equation}}
\def\eeq{\end{equation}}
\def\bsp{\begin{split}}
\def\esp{\end{split}}
\def\bea{\begin{eqnarray}}
\def\eea{\end{eqnarray}}
\def\ba{\begin{array}}
\def\ea{\end{array}}

\def\lb{\left(}
\def\rb{\right)}

\def\l.{\left.}
\def\r.{\right.}

\def\ra{\rangle}
\def\la{\langle}

\def\bo{\bold{k}}

\begin{document}
\author{S. A. Owerre}

\title{Topological honeycomb magnon Hall effect: A calculation of thermal Hall conductivity of magnetic spin excitations}
\affiliation{ African Institute for Mathematical Sciences, 6 Melrose Road, Muizenberg, Cape Town 7945, South Africa.}
\email{solomon@aims.ac.za}
\affiliation{ Perimeter Institute for Theoretical Physics, 31 Caroline St. N., Waterloo, Ontario N2L 2Y5, Canada.}

\begin{abstract}
Quite recently, magnon Hall effect of spin excitations   has been observed experimentally on the kagome and pyrochlore lattices. Thermal Hall conductivity $\kappa^{xy}$, changes sign as a function of magnetic field or temperature on the kagome lattice, and $\kappa^{xy}$  changes sign upon reversing the sign of the  magnetic field on the pyrochlore lattice. Motivated by these recent exciting experimental observations, we theoretically propose a simple realization  of magnon Hall effect in a two-band model on the honeycomb lattice. The magnon Hall effect of spin excitations  arises in the usual way via the breaking of inversion symmetry of the lattice, however, by a  next-nearest-neighbour  Dzyaloshinsky-Moriya (DM) interaction. We find that  $\kappa^{xy}$ has a fixed sign for all parameter regimes considered.   These results are in contrast to the Lieb, kagome and pyrochlore lattices. We further show that the low-temperature dependence on the magnon Hall conductivity follows a $T^{2}$ law, as opposed to the kagome and pyrochlore lattices.  These results suggest an experimental procedure to measure thermal Hall conductivity  within a class of 2D honeycomb quantum magnets and ultracold atoms trapped in honeycomb optical lattice.
\end{abstract}

\maketitle

\textit{ Introduction}.--
 In recent years, the understanding of the topological nature of phonons  and magnons in quantum materials  has been at the pinnacle of intense investigation. These materials are believed to be applicable to many technological systems such as thermal devices and spintronics.  The most fascinating property of these materials is the observation of thermal Hall effect, which occurs at finite temperature. Phonon Hall effect has been observed experimentally in Tb$_3$Ga$_5$O$_{12}$ \cite{stro}, and the topological properties have been studied in terms of the Berry curvature of the system in different lattice geometries \cite{stro1, stro2}.  Recently, the topological properties of magnons in quantum magnets \cite{alex0, alex1, alex2,shin,shin1} have become a subject of interest because of the possibility of thermal Hall effect characterized by a nonzero thermal Hall conductivity  $\kappa^{xy}$, at finite temperature. Thermal Hall effect  in quantum magnets was first predicted theoretically by  Katsura-Nagaosa-Lee \cite{alex0} on the kagome and pyrochlore ferromagnets with a nearest-neighbour (NN) Dzyaloshinsky-Moriya (DM) interaction \cite{dm}. It was later  discovered experimentally  by Onose  {\it et al}   \cite{alex1} in the ferromagnetic insulator Lu$_2$V$_2$O$_7$ on three-dimensional (3D) pyrochlore lattice. Subsequently,  Matsumoto and Murakami \cite{alex2} relates  $\kappa^{xy}$ directly to the Berry curvature of the magnon bulk bands reminiscent of Hall conductivity in electronic systems \cite{thou}.  This result shows that at nonzero temperature,  $\kappa^{xy}\neq 0$  provided that the magnon bulk bands have a nontrivial gap  at the Dirac points, {\it i.e.,}  the points where two bands touch in the Brillouin zone. 

It has also been  shown that $\kappa^{xy}$ changes sign as a function of temperature or magnetic field \cite{alex4a,alex4}.  Also  recently, thermal Hall effect has been observed experimentally  on the 2D kagome magnet Cu(1-3, bdc) \cite{alex6}.  Thus, opening a wide spectrum of research to search for topological magnon insulators in 2D quantum magnets.  In both 2D kagome magnet and 3D pyrochlore magnet, the breaking of inversion symmetry by a NN DM interaction plays a crucial role. The DM interaction introduces a spin-orbit coupling, as a result the bosons accumulate a phase as they hop through the lattice sites, reminiscent of Haldane model in electronic systems \cite{fdm}.  Thus far, thermal Hall effect in quantum magnets has been studied only in three lattice geometries --- the three magnon bulk bands of kagome lattice \cite{alex0,alex6, alex4, alex4a},  the four magnon bulk bands of  pyrochlore lattice \cite{alex1,alex2} and three magnon bulk bands of Lieb lattice \cite{xc}.  However, a topological magnon insulator is also realizable in a two-band honeycomb ferromagnetic model \cite{sol}.  

The purpose of this paper is to provide a simple theoretical realization of magnon Hall effect in this two-band model on the honeycomb lattice. In this system, a nontrivial magnon bulk gap arises in the presence of a next-nearest-neighbour (NNN) DM interaction as opposed to the kagome and the pyrochlore lattices \cite {alex0, alex1, alex2,  alex4, alex6}.   We explicitly demonstrate the theory of thermal Hall effect in this system.  We find that  $\kappa^{xy}$ in this system has a fixed sign for all values of the parameters considered at finite temperatures.   In contrast to 2D kagome and 3D pyrochlore lattices, we  also find that  the low-temperature dependence on the magnon Hall conductivity is $\propto T^{2}$. The model studied here  can be mapped to hardcore bosons on the honeycomb lattice with Haldane-like NNN hopping. In fact, a recent study has reported an experimental realization of the Haldane model  using ultracold fermionic atoms in a periodically modulated optical honeycomb lattice \cite{jot}.  Thus, the results obtained in this paper suggest an experimental procedure to search  a bosonic analogue of Haldane model in  2D honeycomb quantum magnets and ultracold atoms trapped in honeycomb optical lattice.

\begin{figure}[ht]
\centering
\includegraphics[width=3in]{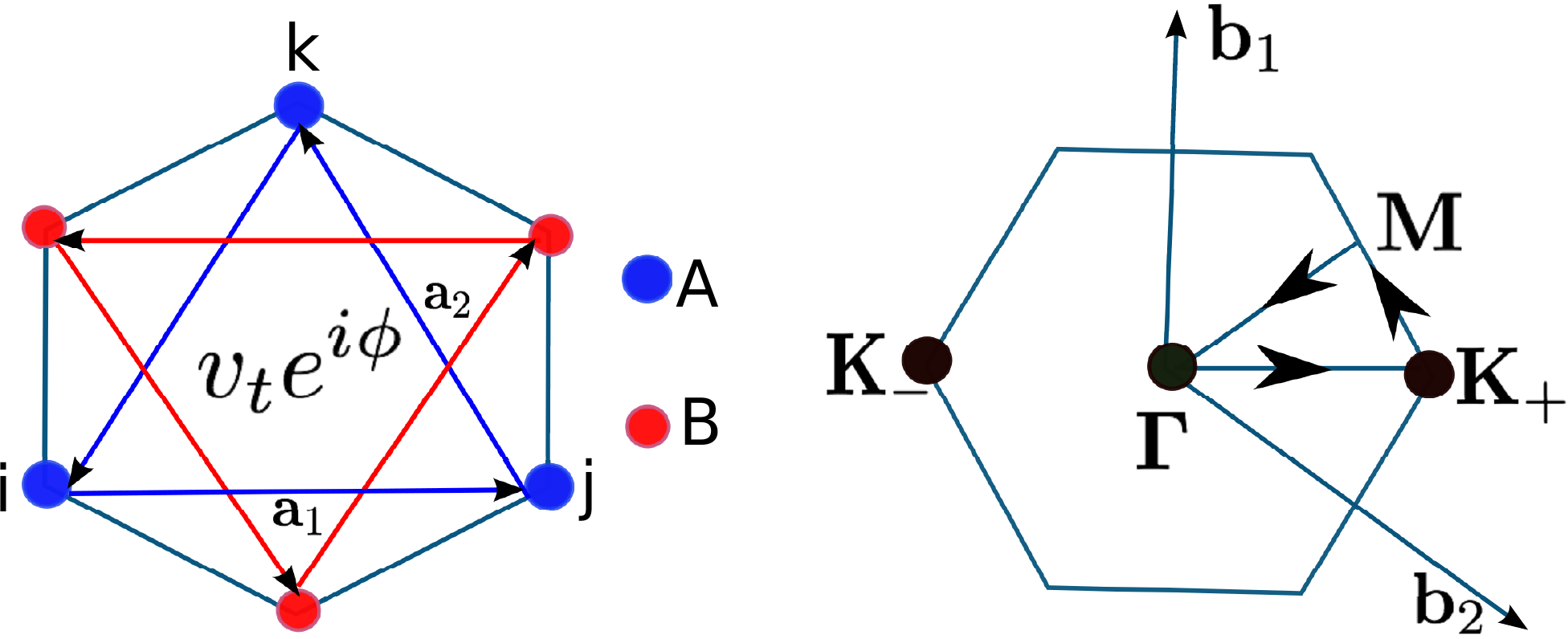}
\caption{Color online. (Left). The unit cell of the honeycomb lattice. The primitive lattice vectors are $\bold a_1=\sqrt{3}a\hat x;~ \bold a_2=a(-\sqrt{3}\hat x, {3}\hat y)/2$.  The arrows show the treads of the magnetic flux generated by the  DM interaction, $\bold{D}_{ij}=\nu_{ij}\bold{ D}\cdot\hat z$, $v_t=S\sqrt{J^{\prime 2} +D^2}$, and $\phi=\arctan(D/J^\prime)$. (Right). The Brillouin zone with reciprocal lattice vectors ${\bf b}_1= 4\pi /3a\hat y$ and ${\bf b}_2= 2\pi(\hat x,-\hat {y}/\sqrt{3})/\sqrt{3}a$.}
\label{unit}
\end{figure}

\textit{Two-band honeycomb ferromagnetic model}.--
 In a recent study, the author has shown the first existence of topological magnon insulator on the honeycomb lattice with two-magnon bands \cite{sol}. In this paper, we provide the first evidence of thermal Hall effect of magnetic spin excitations in this model,  governed by the Hamiltonian \cite{sol}

\begin{align}
H&=-J\sum_{\la ij\ra}{\bf S}_{i}\cdot{\bf S}_{j}-J^\prime\sum_{\la \la ij\ra\ra}{\bf S}_{i}\cdot{\bf S}_{j}+\sum_{\la \la ij\ra\ra} {\bf D}_{ij}\cdot{\bf S}_{i}\times{\bf S}_{j} \nonumber\\&- h\sum_i S_i^z.
\label{h}
\end{align}
The first two terms are ferromagnetic Heisenberg exchange couplings   on the NN sites $J>0$ and NNN sites $J^\prime>0$ respectively. The third term is a NNN DM interaction  between sites $i$ and $j$. The DM interaction generates a magnetic flux. Figure \ref{unit} shows the  direction of the magnetic flux generated by the DM interaction  on the unit cells with $\bold{D}_{ij}=\nu_{ij}\bold{ D}\cdot\hat z$, where  $\nu_{ij}=\pm 1$ for hopping from left to right and vice versa. Thus, the NNN DM interaction plays the role  of  spin-orbit coupling and breaks the inversion symmetry of the lattice.   The last term is the Zeeman magnetic field with $h=g\mu_B B$ being the magnetic field strength, $g$ is the spin $g$-factor and $\mu_B$ is the Bohr magneton. Notice that this system is a ferromagnetic insulator despite the presence of the DM interaction.   
\begin{figure}
\centering
\includegraphics[width=1\linewidth]{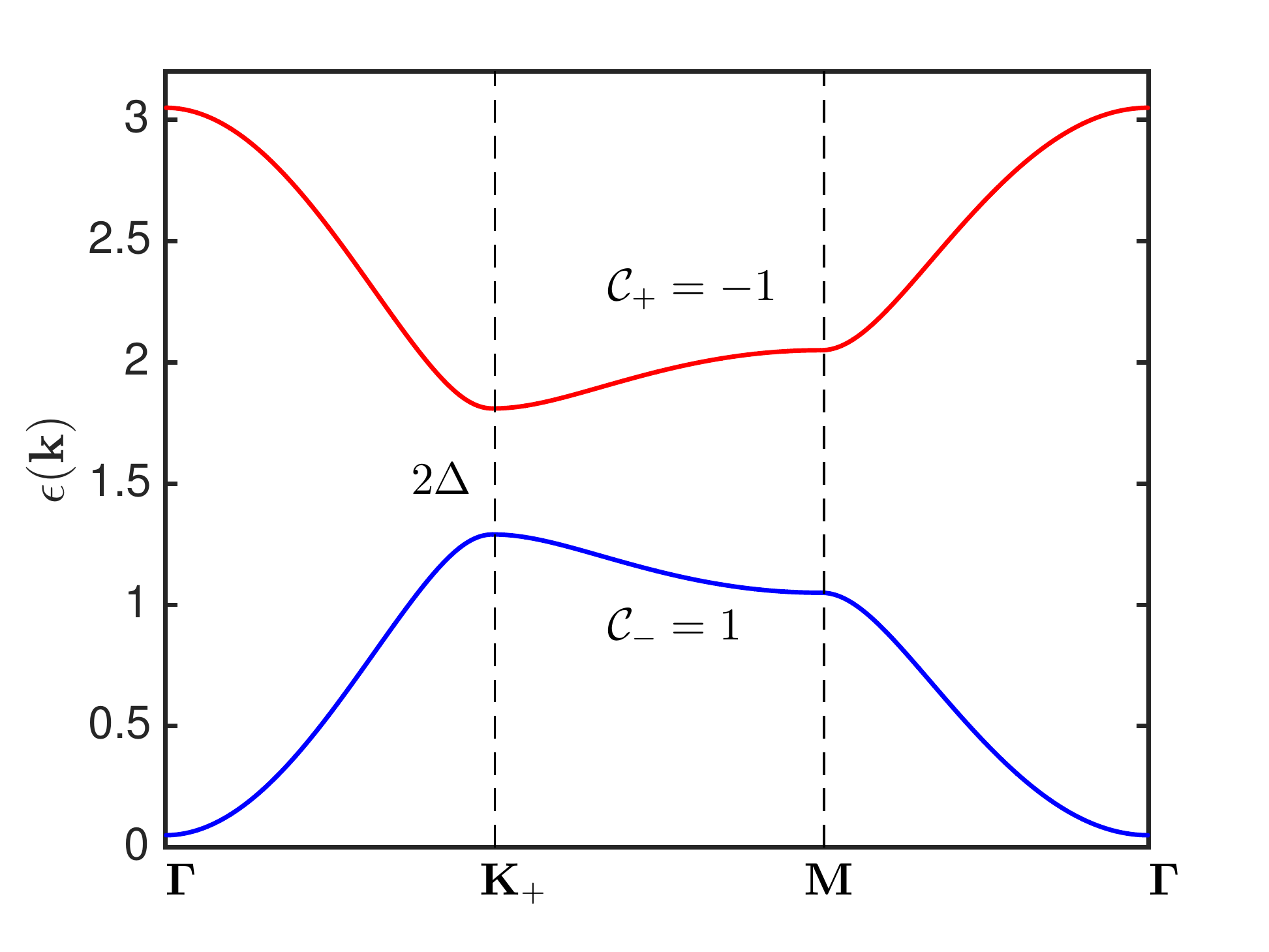}
\caption{Color online. Magnon bulk bands for the spin $1/2$ honeycomb topological magnon insulator in units of $g=J=1$ and the corresponding Chern numbers (see text for explanation) at  $\mu_B B=0.1 $, $J^\prime=D=0.1J,~\phi=\pi/4$.  The gap at ${\bf K}_\pm$ is $\Delta= |h_z(\bold k)|= 3\sqrt{3}v_t\sin\phi= 3\sqrt{3}v_D$. The NNN isotropic interaction does not contribute to gap opening.  The gap at ${\bf \Gamma}$ is $h$. }
\label{HCF}
\end{figure}

It is well-known that the physics of thermal Hall effect can be understood entirely by semiclassical methods \cite{alex0, alex1, alex2, alex4}.  Recent experimental results  on the kagome ferromagnet \cite{alex6} suggest that the Holstein Primakoff transformation  is a better predictor than the Schwinger boson representation. Using the former method,  the spin Hamiltonian maps to a bosonic tight binding hopping model given by
\begin{align}
H&=-v_s\sum_{\la ij\ra}( b_i^\dagger b_j+ h.c.)  - v_t\sum_{\la \la ij\ra\ra}(e^{i\phi_{ij}} b^\dagger_i b_{j}+h.c.)\label{hp3}\\&\nonumber +v_0\sum_i b_i^\dagger b_i,
\end{align}
where $v_0=(zv_s+z^\prime v_s^\prime + h)$,  $v_t=\sqrt{v_s^{\prime 2} +v_D^2}$, $v_s(v_s^\prime)(v_D)= JS(J^\prime S)( DS)$, and  $z(z^\prime)=3(6)$ are the number of NN and NNN sites respectively. We have assumed  a DM interaction along the $z$-quantization axis. As a result of the DM interaction, the bosons accumulate a phase  $\phi_{ij}=\nu_{ij}\phi$, where $\phi=\arctan(D/J^\prime)$ is a magnetic flux generated by the DM interaction on the NNN sites, similar to the Haldane model with $\nu_{ij}=\pm 1$ as in electronic systems \cite{fdm}.  The   momentum space Hamiltonian is given by $H=\sum_{\bold k}\Psi^\dagger_{\bold k}\cdot \mathcal{H}_B(\bold k)\cdot\Psi_{\bold k},$ where $\Psi^\dagger_{\bold k}= (b_{\bold{k} A}^{\dagger},\thinspace b_{\bold{k} B}^{\dagger})$, and the Bogoliubov Hamiltonian is given by  
\begin{align}
\mathcal{H}_B(\bold k)&=h_0(\bold k)\sigma_0 +h_x(\bold k)\sigma_x + h_y(\bold k)\sigma_y + h_z(\bold k)\sigma_z,
\label{honn}
\end{align}
where $h_0= v_0-2v_t\cos\phi p_\bo$, $h_x=-v_s\sum_\mu\cos \bo\cdot\boldsymbol{\delta}_\mu$, $h_y=-v_s\sum_\mu\sin \bo\cdot\boldsymbol{\delta}_\mu$, and $h_z=2v_t\sin\phi \rho_\bo$, where $p_\bo=\sum_\mu\cos \bo\cdot\boldsymbol{a}_\mu$  and $\rho_\bo=\sum_\mu\sin \bo\cdot\boldsymbol{a}_\mu$,
and $ \boldsymbol{\delta}_\mu$ are the three NN vectors on the honeycomb lattice given by $ \boldsymbol{\delta}_1=a(\sqrt{3}\hat x,~\hat y)/2$, $ \boldsymbol{\delta}_2=a(-\sqrt{3}\hat x,\hat y)/2$ and $ \boldsymbol{\delta}_3=a(0, -\hat y)$.  The NNN vectors are shown in Fig.~\ref{unit}.   
 The  corresponding eigenvalues of Eq.~\ref{honn} are given by
 \begin{align}
 \epsilon_{\lambda} (\bold k)&= h_0(\bold k)+\lambda\sqrt{h_x(\bold k)^2 +h_y(\bold k)^2+h_z(\bold k)^2} \label{bandd}\\&\nonumber=h_0(\bold k)+ \lambda\epsilon(\bold k), 
\end{align}
where $\lambda=\pm$ labels the top and the bottom bands respectively.  The explicit form of the eigenvectors  is given by


\begin{equation} \ket{\psi_{\lambda}(\bold{k})} = \frac{1}{\sqrt{2}}\lb\sqrt{1+\lambda\frac{h_z(\bold k)}{\epsilon(\bold k)}}, -\lambda e^{-i\Phi({\bf k})}\sqrt{1-\lambda\frac{h_z(\bold k)}{\epsilon(\bold k)}}\rb^T\label{egv},\end{equation} where 
 \begin{equation}\tan\Phi({\bf k})= \frac{h_y(\bold k)}{h_x(\bold k)}.\end{equation}
 As previously shown \cite{sol, jf},  for $v_D=0$ the system is gapless at two inequivalent Dirac points $\bold{K}_\pm=\pm \frac{4\pi}{3\sqrt 3 a}\hat x$. The NNN interaction $J^\prime$ and the magnetic field $h$ only shift the energy of the Dirac points but do not open a gap.  The gapless Dirac points have been shown to be robust against higher order magnon-magnon interactions \cite{jf}.   As shown  in Fig.~\ref{HCF}, a gap only opens when $v_D\neq 0$, thus gives rise to nonzero Chern numbers defined below.
 
\textit{Berry curvature and Chern number}.--
In fermionic systems, nontrivial band topology  arises from the structure of the energy bands. This is characterized by a nonzero Berry curvature  defined via the eigenstates of the system,  which gives rise to a quantized integer (when the Fermi energy lies between the energy gap) called the Chern number. In bosonic systems, the idea is essentially  the same. A nontrivial band topology arises only when the system exhibits a nontrivial gap in the spin wave excitation spectra and a nonzero Chern number simply predicts the existence of edge state modes in the system.   In 2D lattices, the Berry curvature is generally  given by
\begin{equation}
\Omega_\lambda(\bold k)= -2 \textrm{Im}[\braket{\partial_{k_x}\psi_\lambda(\bold k)|\partial_{k_y}\psi_\lambda(\bold k)}].
\end{equation}
In the present model, the  eigenstates are given by Eq.~\ref{egv}, hence  we obtain
\begin{align}
\Omega_\lambda(\bold k)&=\frac{1}{2}\lambda\epsilon_{\mu\nu}\partial_{k_\mu}\phi(\bold k)\partial_{k_\nu}\lb\frac{h_z(\bold k)}{\epsilon(\bold k)}\rb\label{beryy},
\end{align}
where $\epsilon_{\mu\nu}$ is a 2D antisymmetric tensor. 
\begin{figure*}
\centering
  \subfigure[\label{33a}]{\includegraphics[width=.45\linewidth]{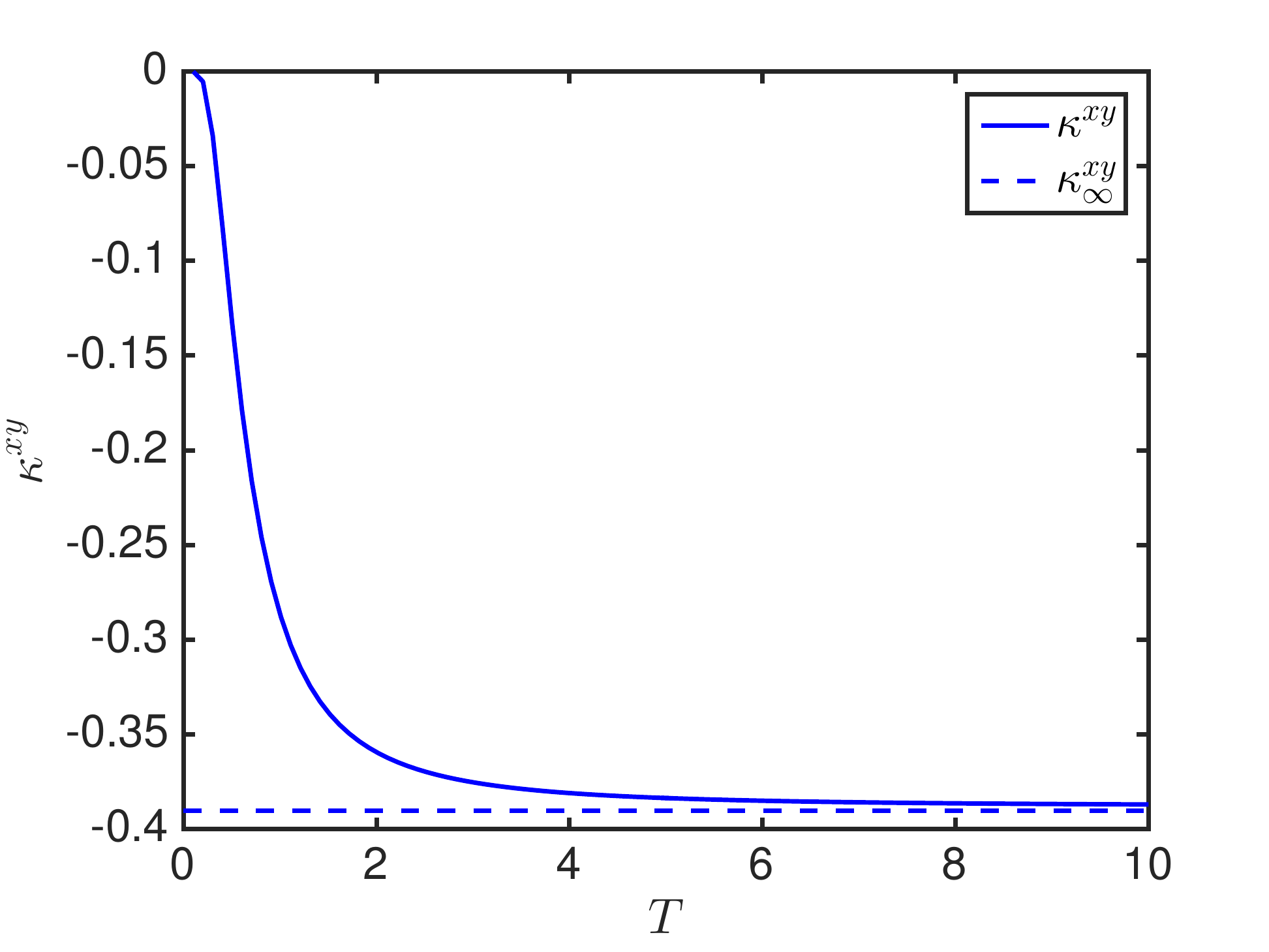}}
   \quad
   \subfigure[\label{33b}]{\includegraphics[width=.45\linewidth]{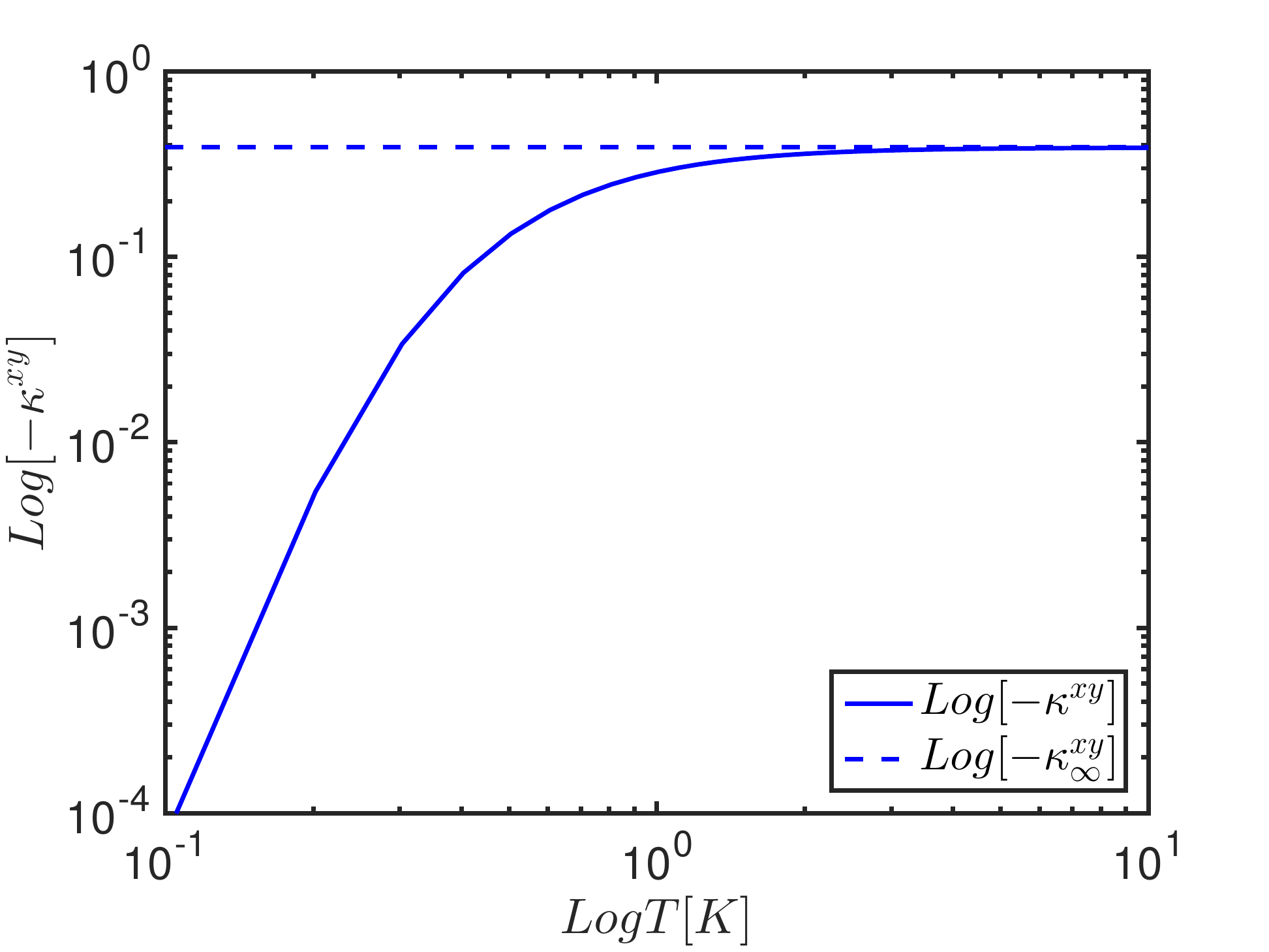}}
\caption{Color online.  The plot of $(a)$ $T$ vs. $\kappa^{xy}$;  $(b)$ $\log[T]$ vs. $\log[-\kappa^{xy}]$ with  the same parameters in Fig.~\ref{HCF}.}
\end{figure*}

\begin{figure*}
\centering
  \subfigure[\label{3a}]{\includegraphics[width=.45\linewidth]{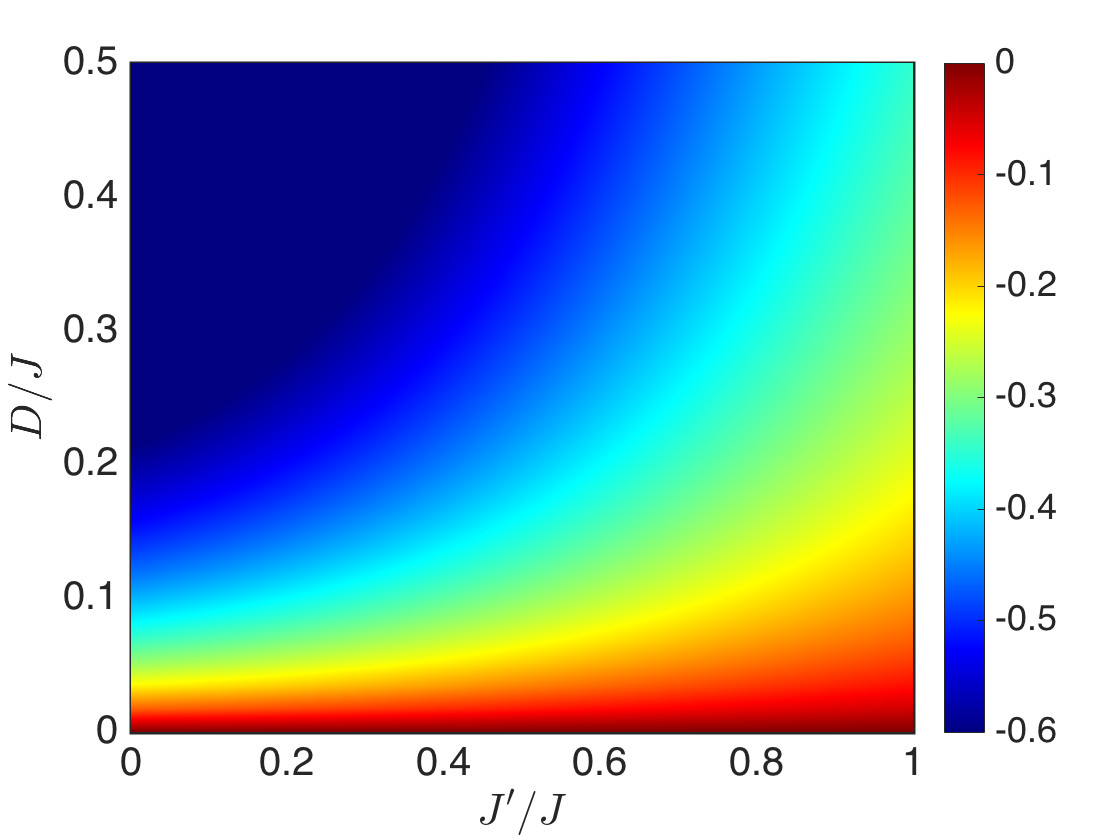}}
   \quad
   \subfigure[\label{3b}]{\includegraphics[width=.45\linewidth]{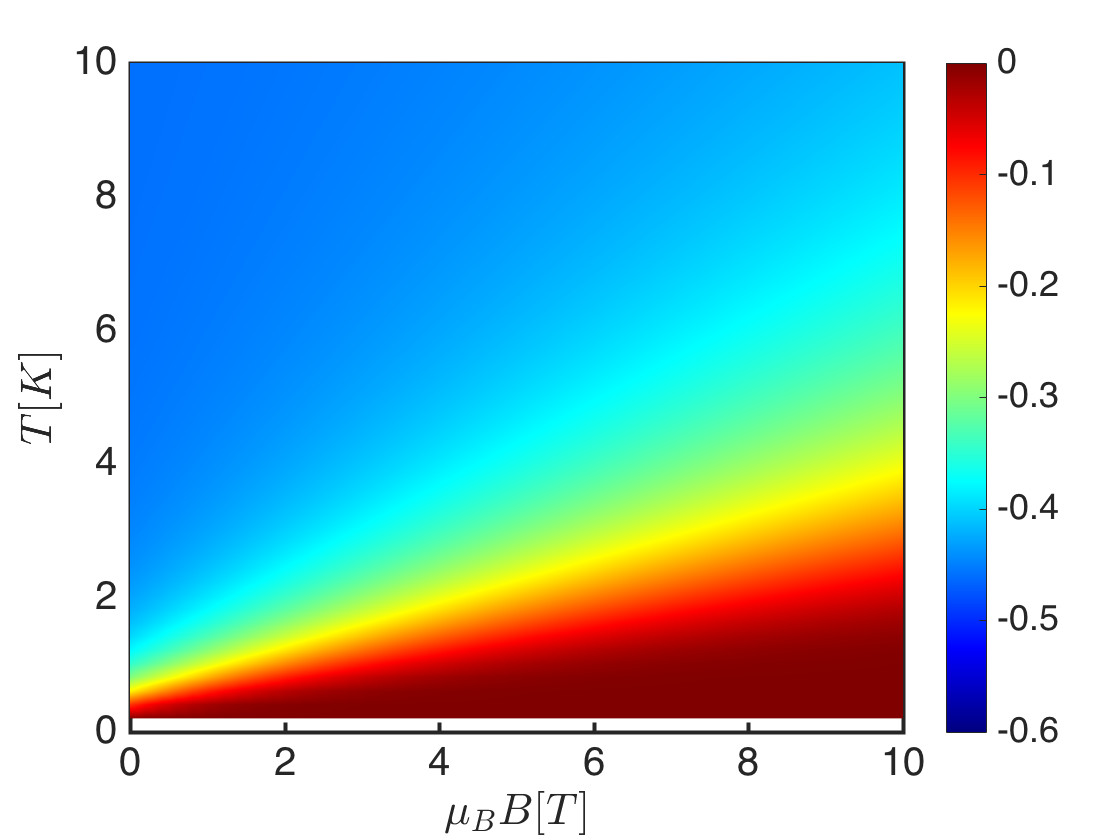}}
\caption{Color online. The contour plot of $\kappa^{xy}$ in the $(a)$ $D/J$ vs. $J^\prime/J$ plane at $T=1.5$ and $\mu_B B=0.1$.  $(b)$  $\mu_B B$ vs. $T$ plane with the parameters  in Fig.~\ref{HCF}.}
\end{figure*}
 The Berry curvature has two important properties. Firstly, it  is controlled by the mass gap at the Dirac points and the Chern number
\begin{equation}
\mathcal{C}_\lambda= \frac{1}{2\pi}\int_{{BZ}} d^2k~ \Omega_\lambda(\bold k),
\label{chenn}
\end{equation}
of a gapless system $v_D=0$ vanishes. Secondly,  for a gapped system $v_D\neq 0$,  the maximum contribution to the Chern number comes from the states near ${\bf K}_\pm$, {\it i.e.} when $\Delta= |h_z(\bold k)|\to 0$, also the total Chern number vanishes. Therefore, no magnon edge state exists above the upper band \cite{sol}.

\textit{Magnon Hall effect}.--
The nontrivial topology of the Berry curvatures lead to magnon edge states \cite{sol}.  Interestingly, the magnon edge states  carry a transverse heat (spin) current upon the application of a longitudinal temperature gradient. As magnons are uncharged particles, there is no Lorentz force,  the DM interaction plays the role of an effective magnetic field by altering the propagation of the magnon in the system, thus leads to  magnon Hall effect \cite{alex1}.   In the following, we demonstrate how the topological nature of this system is manifested by computing important experimental observables.  The important quantity characterizing the magnon Hall effect is the thermal Hall conductivity. Similar to Hall conductivity in electronic systems, thermal Hall conductivity  is related to the Berry curvature of the eigenstates.  There are two contributions to the thermal conductivity given by \cite{alex2}
 \begin{align}
\kappa_E^{xy}&=-\frac{2 }{T}\text{Im}\sum_{\lambda=\pm}\int_{{BZ}} \frac{d^2k}{(2\pi)^2} n_B[\epsilon_\lambda(\bold k)]\nonumber\\&\times [\braket{\partial_{k_x}\psi_\lambda(\bold k)|\lb \mathcal{H}_B(\bold k)+\epsilon_\lambda(\bold k)\rb^2|\partial_{k_y}\psi_\lambda(\bold k)}],
\label{thm1}
\end{align}
and
\begin{align}
\kappa_O^{xy}=-\kappa_E^{xy}-\frac{k_B^2 T}{(2\pi)^2}\sum_{\lambda=\pm}\int_{{BZ}} d^2k c_2\lb n_\lambda\rb\Omega_\lambda(\bold k),
\label{thm2}
\end{align}
where
$n_\lambda\equiv n_B[\epsilon_\lambda(\bold k)]=[e^{{\epsilon_\lambda(\bold k)}/k_BT}-1]^{-1}$ is the Bose function, $c_2(x)=(1+x)\lb \ln \frac{1+x}{x}\rb^2-(\ln x)^2-2\textrm{Li}_2(-x),$ and $\text{Li}_n(x)$ is a polylogarithm.  The first contribution originates from the current density \cite{alex0} and the second contribution stems from the orbital motions of magnons \cite{alex2}. Hence, the total contribution is given by
\begin{align}
\kappa^{xy}=\kappa_E^{xy}+\kappa_O^{xy}=-\frac{k_B^2 T}{(2\pi)^2}\sum_{\lambda=\pm}\int_{{BZ}} d^2k c_2\lb n_\lambda\rb\Omega_\lambda(\bold k).
\label{thm}
\end{align}
  It is evident that the thermal Hall conductivity is the Berry curvature weighed by the $c_2(n_\lambda)$ function. It  depends on  the existence of a nontrivial gap in the magnon bulk bands  at $T\neq 0$.   The largest contribution to $\kappa^{xy}$  comes from the states near ${\bf K}_\pm$ due to the Berry curvature as mentioned above. 
In what follows, we apply Eq.~\ref{thm} to the two-band model and  we assume units such that $J=k_B=\hbar=1$.
\begin{figure}
\centering
\includegraphics[width=.75\linewidth]{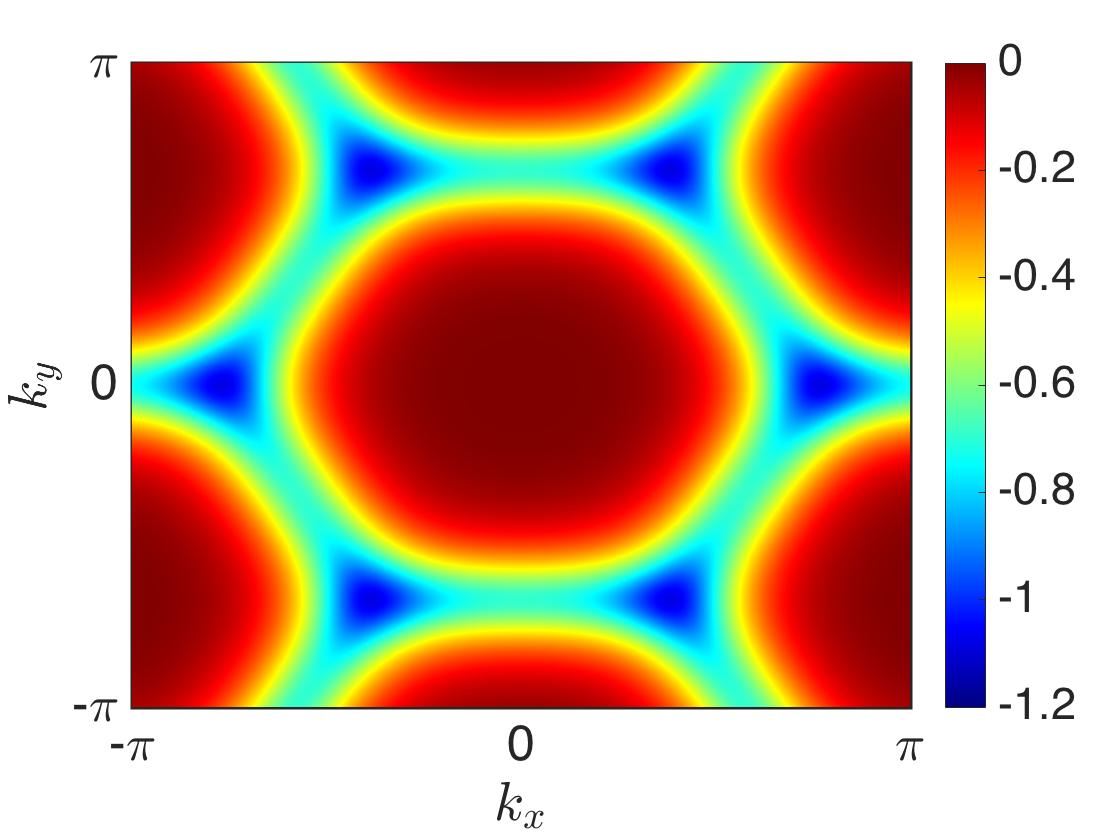}
\caption{Color online. The integrand $\sum_{\lambda=\pm} \epsilon_\lambda\Omega_\lambda(\bold k)$ in Eq.~\ref{high1} describing the high-temperature limit of the thermal Hall conductivity.  The parameters are the same as in Fig.~\ref{HCF}.}
\label{Berry}
\end{figure} 

 Figures~\ref{33a} and \ref{33b} show the plot of  $T$ vs. $\kappa^{xy}(T)$ and $\log T$ vs. $\log[-\kappa^{xy}(T)]$  respectively,  and  Figs.~\ref{3a} and \ref{3b} show the contour plots of $\kappa^{xy}$ in  $J^\prime/J$ vs. $D/J$  plane  and $\mu_B B$ vs. $T$ plane respectively. We observe no sign change in $\kappa^{xy}$ for all   parameter regimes considered. This can be understood directly from Eq.~\ref{thm}. Since the Berry curvatures are equal and opposite $\Omega_+=-\Omega_-$, Eq.~\ref{thm} involves the difference between the $c_2[n(\epsilon_\lambda, T)]$ functions. At low temperatures, the lower band dominates, but due to the fact that  $c_2[n(\epsilon_-, T)]>0$ and $\Omega_- >0$   for all parameter regimes,  the thermal Hall conductivity is negative, $\kappa^{xy}<0$, in the low-$T$ limit. At high temperatures,  the $T\to \infty$ limit of $\kappa^{xy}(T)$ is given by \cite{alex4}
\begin{align}
\kappa^{xy}_\infty= \frac{k_B^2}{(2\pi)^2}\sum_{\lambda=\pm}\int_{{BZ}} d^2k \epsilon_\lambda\Omega_\lambda(\bold k).
\label{high1}
\end{align} 
In Fig.~\ref{Berry}, we show the integrand in Eq.~\ref{high1} describing the high-temperature limit of the thermal Hall conductivity. As mentioned above, it is obvious that the dominant  contribution comes from   the Dirac points $\bold K_\pm$. Because of the fact that $\epsilon_+>\epsilon_-$, the integrand is dominated by the upper band, but since $\Omega_+=-\Omega_- <0$  the integrand is negative. Again  $\kappa^{xy}_\infty<0$ as shown in Figs.~\ref{33a} and \ref{33b}. Therefore, the high and low temperature limits of $\kappa^{xy}(T)$ have the same sign. The sign of $\kappa^{xy}$ is inherited from  the topology of the magnon bulk bands.  For the kagome lattice, Ref.~\cite{alex4a} explains the sign change in $\kappa^{xy}$  as a consequence of the sign change in  Berry curvature of the highest band and   Ref.~\cite{alex4} argues that the sign change in $\kappa^{xy}$ is a consequence of the propagation of the magnon edge states, however, with a NNN interaction. The origin of the sign change on the  kagome \cite{alex0, alex4, alex4a} and Lieb \cite{xc} ferromagnets is still not well-understood theoretically.    Since both  lattices have a flat band, it  may be possible that this flat band has an effect on  the nature of the thermal conductivity. 

Next, we calculate the explicit form of the low-temperature dependence of Eq.~\ref{thm1}. As mentioned above, the lower band dominates at low temperature. Inserting a complete set of states into Eq.~\ref{thm1} gives \cite{alex1}
 \begin{align}
\kappa_E^{xy} &\sim \frac{1}{T}\int \frac{d^2k}{(2\pi)^2} n_B[\epsilon_-(\bold k)]\lb\epsilon_-(\bold k)+\epsilon_+(\bold k)\rb^2\Omega_-(\bold k).
\label{thm4}
\end{align}
The dominant contribution comes from the $\bold k =0$ mode.   Near  $\bold k=0$, $\epsilon_-(\bold k)+\epsilon_+(\bold k)\sim 2(h +3v_s)$ and   $\epsilon_-(\bold k)\sim h +\frac{3v_s}{4} k^2$. Since the Berry curvature $\Omega_-(\bold k)$ vanishes at $\bold k=0$, we must expand it to quadratic order. The  denominator  is finite at  $\bold k=0$, so we expand the numerator as  $\bold k\to 0$.  The only nonzero term  in the expansion up to quadratic order is given by
\begin{align}
\Omega_-(\bold k)\sim \alpha k_x^2k_y^2,
\end{align}
where $\alpha= 3\sqrt{3}v_t\sin\phi/32v_s$. Performing the angular part of the integration,  Eq.~\ref{thm4} takes the form
 \begin{align}
\kappa_E^{xy} &\sim\frac{ \alpha(h+3v_s)^2}{4\pi T}\int_0^\infty k ~dk \frac{k^4}{e^{(h + \frac{3v_s}{4} k^2)/T}-1},\nonumber\\&=\lb \frac{T}{v_s^2}\rb ^{2}\frac{v_t\sin\phi(h+3v_s)^2}{6\sqrt{3}\pi}\text{Li}_{3}\bigg[ \exp{\lb-\frac{h}{T}\rb}\bigg].
\label{thm5}
\end{align}
 This expression  [Eq.~\ref{thm5}] shows that the  honeycomb chiral ferromagnet has a different feature from the kagome and pyrochlore lattices \cite{alex1, alex0}.

\textit{Conclusion}.-- The main result of this paper is that the magnon Hall effect is realizable in a two-band model on the honeycomb lattice. In this paper, we have shown that the thermal Hall conductivity on the honeycomb lattice does not change sign as the temperature or magnetic field vary, with other parameters  fixed.  We also showed that the low-temperature dependence on the thermal Hall conductivity is also different from the kagome and pyrochlore lattices and follows the relation: $\kappa_E^{xy}\propto T^{2}$.  As previously mentioned, the model studied is related to hard-core bosons, and the resulting Hamiltonian is analogous to Haldane model in electronic systems.  Hence,  magnon Hall effect on the honeycomb lattice could be   accessible  using  ultracold bosonic atoms trapped in honeycomb optical lattice. Although honeycomb ferromagnets are difficult to find in nature, they exist in intra-layer regions in some materials and can be considered as two-dimensional materials if the intra-layer coupling, say $J_t$ is larger than the inter-layer coupling, say $J_l$. This is the case in the crystal CrBr$_3$, which comprises strong nearest-neighbour honeycomb ferromagnetic intra-layer coupling \cite{dav0,dav}.  It might be possible to induce a DM interaction in this ferromagnetic material and the results of this paper can be confirmed directly. This might also be applicable to the field of spintronics.

\textit{Acknowledgments}.--
 The author would like to thank African Institute for Mathematical Sciences (AIMS). Research at Perimeter Institute is supported by the Government of Canada through Industry Canada and by the Province of Ontario through the Ministry of Research
and Innovation.


\begin{thebibliography}{0}
\bibitem{stro} C. Strohm, G. L. J. A. Rikken, and P. Wyder, Phys. Rev. Lett.,  {\bf 95}, (2005) 155901.

\bibitem{stro1} { L. Zhang, J. Ren, J.-S. Wang, and B. Li} {Phys. Rev. Lett.}  {\bf 105}, {(2010)} {225901}.
\bibitem{stro2}    {L. Zhang, J. Ren, J.-S. Wang, and B. Li} {J. Phys. Condens. Matter}{\bf 23}, {(2011)} {305402}.
\bibitem{alex0}
{H. Katsura, N. Nagaosa, and P. A. Lee} {Phys. Rev. Lett.} {\bf 104}, {(2010)} {066403}.
\bibitem{alex1}
  {S.  Y. Onose}  {\it et al}~   {Science}  { \bf 329}, {(2010)} {297}.
 \bibitem{alex2}
  {R. Matsumoto and S. Murakami} {Phys. Rev. Lett.}  {\bf 106}, {(2011)} {197202}; {Phys. Rev. B} {\bf 84}, {(2011)} {184406}.
\bibitem{shin}
 R. Shindou et.al., Phys. Rev. B 87,  174427 (2013); Phys. Rev. B 87, 174402 (2013).
  \bibitem{shin1}
  R. Matsumoto, R. Shindou, and S. Murakami, Phys. Rev. B 89, 054420 (2014).
   \bibitem{dm}
 { I. Dzyaloshinsky}  {J. Phys. Chem. Solids} {\bf 4}{1958}{241}; {T. Moriya} {Phys. Rev. }{\bf 120}, {(1960)} {91}.
  
 \bibitem{thou}
 {D. J. Thouless, M. Kohmoto, M. P. Nightingale, and M. den Nijs} {Phys. Rev. Lett. } {\bf 49}, {(1982)} {405}; M. Kohmoto {Annals of Physics} {\bf 160}, {(1985)} {343}.
 \bibitem{alex4a}
 { H. Lee, J. H. Han, and P. A. Lee}   {Phys. Rev. B.} {\bf 91}, {(2015)} {125413}.
\bibitem{alex4}
 {A.  Mook, J.  Henk, and I. Mertig} {Phys. Rev. B} {\bf 90}, {(2014)}{024412}; A.  Mook, J.  Henk, and I. Mertig,  {Phys. Rev. B} {\bf 89}, {(2014)} {134409}.

 \bibitem{alex6}
 {M.  Hirschberger} {\it et al}~  {Phys. Rev. Lett.}  {\bf 115}, {(2015)}{106603}.
\bibitem{fdm}
 {F. D. M. Haldane} {Phys. Rev. Lett.} {\bf 61}, {(1988)}{2015}.
  \bibitem{xc} {X. Cao, K. Chen and D.  He}, {J. Phys.: Condens. Matter} {\bf 27}, {(2015)} {166003}.

\bibitem{sol}
 S. A.  Owerre, J. Phys.: Condens. Matter 28, (2016) 386001.

\bibitem{jot}
 {G. Jotzu} {\it et al},~ Nature {\bf 515}, {(2014)} {237}.


 \bibitem{jf}
  {J. Fransson, A. M. Black-Schaffer, A. V. Balatsky}  {arXiv:1512.04902}.
  \bibitem{dav0}
  A. C. Gossard, V. Jaccarino, and J. P. Remeika, Phys. Rev. Lett. {\bf 7}, (1961) 122 
  \bibitem{dav}
H. L. Davis and Albert Narath, Phys. Rev. {\bf 134}, (1964) A433


\end{thebibliography}
\end{document}